\documentstyle[epsf,12pt]{article}

\renewcommand{\baselinestretch}{1.23}

\begin{document}


\def\be{\begin{eqnarray}}
\def\en{\end{eqnarray}}
\def\non{\nonumber}
\def\la{\langle}
\def\ra{\rangle}
\def\up{\uparrow}
\def\down{\downarrow}
\def\ep{\varepsilon}
\def\ums{{\mu}_{_{\overline{\rm MS}}}}
\def\u{\mu_{\rm fact}}
\def\gg{\Delta\sigma^{\gamma G}}
\def\lsim{ {\ \lower-1.2pt\vbox{\hbox{\rlap{$<$}\lower5pt\vbox{\hbox{$\sim$}
}}}\ } }
\def\gsim{ {\ \lower-1.2pt\vbox{\hbox{\rlap{$>$}\lower5pt\vbox{\hbox{$\sim$}
}}}\ } }
\def\pr{{\sl Phys. Rev.}~}
\def\prl{{\sl Phys. Rev. Lett.}~}
\def\pl{{\sl Phys. Lett.}~}
\def\np{{\sl Nucl. Phys.}~}
\def\zp{{\sl Z. Phys.}~}

\title{
\begin{flushright}
{\normalsize IP-ASTP-05-96}
\end{flushright}
\vskip 10mm
\Large\bf
Can $B\to J/\psi K(K^*)$ Decays Be Described by Factorization ?
}
\author{{\bf Hai-Yang Cheng}\\
{\it Institute of Physics, Academia Sinica}\\
{\it Taipei, Taiwan 115, Republic of China}\\
}
\date{(October, 1996)}

\maketitle

\begin{abstract}
  The new measurements of $B\to J/\psi K(K^*)$ decays by CDF and CLEO indicate
that the production ratio $R$ and the fraction of longitudinal polarization 
$\Gamma_L/\Gamma$ are smaller than the previous results. In conjunction
with the new result of parity-odd transverse polarization in $B\to
J/\psi K^*$, we found a minimal modification to the factorization
hypothesis: While the data of $B\to J/\psi K^*$ can be accommodated in the
factorization approach with nonfactorizable terms $
\chi_{_{A_1}}=\chi_{_{A_2}}=\chi_{_{V}}\equiv\chi$ of order 15\%, the
result of $R$ measurement requires that the nonfactorizable effect  
$\chi_{_{F_1}}$ on $B\to J/\psi K$ be slightly larger than $\chi$. 
Therefore, the effective parameter $a_2^{\rm eff}$ is not universal even 
for $B\to J/\psi K(K^*)$ decays.
We have generalized the considerations to $B\to \psi(2S)K(K^*)$ and $B_s\to 
J/\psi\phi$ 
and found that the predictions are in agreement with currently available data.


\end{abstract}
\pagebreak

  The two-body nonleptonic weak decays of $D$ and $B$ mesons are conventionally
described by the factorization approach. It has been shown 
\cite{Gour94,Alek1} that this approach fails to account for the observed
fraction of longitudinal polarization $\Gamma_L/\Gamma$ in $B\to J/\psi K^*$
decays in all commonly used models of form factors and the data of
the production ratio $R\equiv\Gamma(B\to J/\psi K^*)/\Gamma(B\to J/\psi
K)$ in many known models. The issue of how to test or modify the factorization
hypothesis to describe the data of $B\to J/\psi K(K^*)$ has been the
subject of many subsequent studies. Recently, CDF \cite{CDF96} and CLEO 
\cite{CLEO96} have presented new measurements of $B\to J/\psi
K(K^*)$ decays and for the first time CLEO has measured the parity-odd 
transverse polarization in $B\to J/\psi K^*$.
It turns out that the new results of $\Gamma_L/\Gamma$ and $R$ are smaller
than the previous values. The purpose of this Letter is to study the 
theoretical implications of the new data.

   Under the factorization approximation, the hadronic matrix element is
factorized into the product of two matrix elements of single currents,
governed by decay constants and form factors. For the QCD-corrected weak
Hamiltonian
\be
{\cal H}_{\rm eff}\,\propto~c_1O_1+c_2O_2=c_1(\bar{q}_1q_2)(\bar{q}_3q_4)+
c_2(\bar{q}_1q_4)(\bar{q}_3q_2),
\en
where $(\bar{q}_1q_2)\equiv \bar{q}_1\gamma_\mu(1-\gamma_5)q_2$,
the external $W$-emission and internal $W$-emission amplitudes
(or the so-called class-I and class-II decay amplitudes) characterized by
the parameters $a_1,~a_2$ are related to the Wilson coefficient functions
$c_1$ and $c_2$ via
\be
a_1=c_1+{c_2\over N_c},~~~~a_2=c_2+{c_1\over N_c}.
\en
Factorization requires that
$a_1$ and $a_2$ be universal; i.e., they are channel-independent in $D$ or
$B$ decays. However, we have learned from charm decays a long time ago that
the naive factorization approach never works for the decay rate of 
color-suppressed (i.e.,
class-II) decay modes, though it might work for class-I decays \cite{Cheng89}.
The most noticeable example is the ratio $\Gamma(D^0\to\bar{K}^0\pi^0)/
\Gamma(D^0\to K^-\pi^+)$, which is predicted to be $\sim 0.02$ whereas
experimentally it is measured to be $0.55\pm 0.06$ \cite{PDG}.
This implies that the inclusion of nonfactorizable contributions is 
inevitable and necessary. Since the amplitudes of $D,~B\to PP,~VP$ decays
($P$: pseudoscalar meson, $V$: vecetor meson) are governed by a single form
factor, the effects of nonfactorization amount to a redefintion of 
the effective parameters $a_1$ and $a_2$ \cite{Cheng94}:
\be
a_1^{\rm eff}=c_1+c_2\left({1\over N_c}+\chi_1\right),~~~~~a_2^{\rm eff}=c_2
+c_1\left({1\over N_c}+\chi_2\right),
\en
or
\be
a_1^{\rm eff}=a_1\left(1+{c_2\over a_1}\chi_1\right),~~~~~a_2^{\rm eff}=a_2
\left(1+{c_1\over a_2}\chi_2\right).
\en
For the expressions of the nonfactorizable terms denoted by the parameters
$\chi_1$ and $\chi_2$, consider the decay $\bar B\to D\pi$ as an example;
we obtain \cite{Soares,KS95}
\be
\chi_{_1}(\bar B\to D\pi) = \chi_{_{1,F_0^{BD}}}={\la D\pi|\widetilde{O}_1|
\bar B\ra\over \la
D\pi|O_1|\bar B\ra_f} +{a_1\over c_2}\,{\la D\pi|{O}_1|\bar B\ra_{nf}\over 
\la D\pi|O_1|\bar B\ra_f}={F_0^{(8)nf}(m_\pi^2)\over F_0^{BD}(m_\pi^2)}
+{a_1\over c_2}\,{F_0^{(1)nf}(m_\pi^2)\over F_0^{BD}(m_\pi^2)}, &&  \non \\
\chi_{_2}(\bar B\to D\pi) =\chi_{_{2,F_0^{B\pi}}}={\la D\pi|\widetilde{O}_2|
\bar B\ra\over \la
D\pi|O_2|\bar B\ra_f} +{a_2\over c_1}\,{\la D\pi|{O}_2|\bar B\ra_{nf}\over 
\la D\pi|O_1|\bar B\ra_f}={F_0^{(8)nf}(m_D^2)\over F_0^{B\pi}(m_D^2)}
+{a_2\over c_1}\,{F_0^{(1)nf}(m_D^2)\over F_0^{B\pi}(m_D^2)}, &&
\en
where the subscripts $f$ and $nf$ denote factorizable and nonfactorizable
contributions, respectively, and use has been made of the Fierz identify
\be
O_{1,2}=\,{1\over N_c}O_{2,1}+\widetilde{O}_{2,1}
\en
with $\widetilde{O}_1={1\over 2}(\bar{q}_1\lambda^a q_2)(\bar{q}_3\lambda^a
q_4)$ being a product of the color-octet currents $(\bar{q}\lambda^a
q')\equiv\bar{q}\gamma_\mu(1-\gamma_5)\lambda^a q'$. Since the factorizable
$B\to D\pi$ amplitudes are
\be
\la D\pi|O_1|B\ra_f &=& f_\pi(m_B^2-m_D^2)F_0^{BD}(m_\pi^2), \non \\
\la D\pi|O_2|B\ra_f &=& f_D(m_B^2-m_\pi^2)F_0^{B\pi}(m_D^2),
\en
we have followed \cite{KS95} to define the form factors $F_0^{(1)nf}$ and
$F_0^{(8)nf}$:
\be
\la D\pi|O_1|B\ra_{nf} &=& f_\pi(m_B^2-m_D^2)F_0^{(1)nf}(m_\pi^2), \non \\
\la D\pi|\widetilde{O}_1|B\ra &=& f_\pi(m_B^2-m_D^2)F_0^{(8)nf}(m_\pi^2).
\en
Likewise for $\bar{B}\to D\rho$ and $D^*\pi$ decays, we have
\be
&& \chi_{_1}(\bar B\to D\rho)=\chi_{_{1,F_1^{BD}}}\,,~~~~\chi_{_2}(\bar B\to D
\rho)=\chi_{_{2,A_0^{B\rho}}}\,,  \non \\
&& \chi_{_1}(\bar B\to D^*\pi)=\chi_{_{1,A_0^{BD^*}}}\,,~~~~\chi_{_2}(\bar B
\to D^*\pi)=\chi_{_{2,F_1^{B\pi}}}\,.
\en
Since $c_1/a_2>>1$, it is clear from Eqs.(4) and (5) that nonfactorizable 
effects are more dramatic in color-suppressed decay modes. As a consequence,
a determination of $\chi_2$ is more reliable than $\chi_1$.

   The study of nonfactorizable effects in $M\to VV$ decay is more
complicated as its general amplitude consists of three independent
Lorentz scalars:
\be
A[M(p)\to V_1(\ep_1,p_1)V_2(\ep_2,p_2)]\propto \ep^*_\mu(\lambda_1)
\ep^*_\nu(\lambda_2)(\hat{A}_1g^{\mu\nu}+\hat{A}_2p^\mu p^\nu+i
\hat{V}\epsilon^{\mu\nu\alpha\beta}p_{1\alpha}p_{2\beta}),
\en
where $\hat{A}_1,~\hat{A}_2,~\hat{V}$ are related to the form factors
$A_1,~A_2$ and $V$, corresponding to $S$-, $D$- and $P$-waves,
respectively. For $\bar B\to D^*\rho$ we can
define various nonfactorizable terms:
\be
&& \chi_{_{1,A_1}}(\bar B\to D^*\rho)=\chi_{_{1,A_1^{BD^*}}},~~~
\chi_{_{2,A_1}}(\bar B\to D^*\rho)=\chi_{_{2,A_1^{B\rho}}}, \non \\
&& \chi_{_{1,A_2}}(\bar B\to D^*\rho)=\chi_{_{1,A_2^{BD^*}}},~~~
\chi_{_{2,A_2}}(\bar B\to D^*\rho)=\chi_{_{2,A_2^{B\rho}}},  \\
&& \chi_{_{1,V}}(\bar B\to D^*\rho)=\chi_{_{1,V^{BD^*}}},~~~
\chi_{_{2,V}}(\bar B\to D^*\rho)=\chi_{_{2,V^{B\rho}}}. \non 
\en
Since {\it a priori} there is no reason to
expect that nonfactorizable terms weight in the same way to $S$-, $P$-
and $D$-waves, namely $\chi_{_{i,A_1}}=\chi_{_{i,A_2}}=\chi_{_{i,V}}$,
it is in general not possible to
define an effective $a_1$ or $a_2$ for $M\to VV$ decays once nonfactorizable
effects are taken into account \cite{KS94}. However, if nonfactorizable
contributions are channel-independent, i.e.,
\be
\chi_{_{i,F_0}}=\chi_{_{i,F_1}}=\chi_{_{i,A_0}}=\chi_{_{i,A_1}}=
\chi_{_{i,A_2}}=\chi_{_{i,V}}\equiv\chi,
\en
the effective parameters $a_1^{\rm eff}$ and $a_2^{\rm eff}$ given by Eq.(3)
or (4) are applicable not only to $M\to PP,~VP$ but also to $M\to VV$ and 
they become universal. In this case we have a new
factorization scheme with $\chi\neq 0$. Note that the predictions for the
ratios of physical quantities (e.g., $\Gamma_L/\Gamma$ and $|P|^2$)
in the same class of decay modes in the new
factorization method are the same as that in naive factorization 
(i.e., $\chi=0$). Empirically, it has been found that 
that the discrepancy between theory and experiment for charm decays is greatly
improved if Fierz transformed terms in (3) are dropped \cite{Fuk,Cheng89}. It 
has been argued that this empirical observation is justified in the 
so-called large-$N_c$ approach in which a rule of discarding subleading 
$1/N_c$ terms can be formulated \cite{Buras}. This amounts to having 
universal nonfactorizable terms $\chi_1=\chi_2=-1/N_c$ in Eq.(3). 
In reality, there is no reason to expect a universal $\chi_1$ or 
$\chi_2$. Since $\chi_2$ comes mainly from color-octet currents, it is natural
to expect that \cite{Cheng94,Cheng95}
\be
|\chi_2(D\to VP)|\gsim|\chi_2(D\to PP)|>|\chi_2(B\to VP)|,
\en
as nonfactorizable soft-gluon effects become more important when final 
particles move slower, allowing more time for significant final-state 
interactions. Indeed, the above theoretical expectation is confirmed by
the phenomenological analyses of $D$ and $B$ decay data \cite{Cheng94,Cheng95}.
For charm decay, $\chi_2$ is found to be in the range $-0.60<\chi_2(D)<-1/3$
\cite{Cheng95}.

   We now come back to $B\to J/\psi K(K^*)$ decays. The general
expressions for the fraction of longitudinal polarization $\Gamma_L/\Gamma$ 
and production ratio $R$ read
\be
{\Gamma_L\over \Gamma} &\equiv& {\Gamma(B\to J/\psi K^*)_L\over \Gamma(B\to 
J/\psi K^*)}=\left[a\left(1+{c_1\over a_2}\chi_{_{A_1}}\right)-bx
\left(1+{c_1\over a_2}\chi_{_{A_2}}\right)\right]^2/\xi, \non \\
R &\equiv& {\Gamma(B\to J/\psi K^*)\over \Gamma(B\to J/\psi K)} =\,1.08\,
{\xi\over z^2\left(1+{c_1\over a_2}\chi_{_{F_1}}\right)^2 },
\en
where
\be
&& \xi=\left[a\left(1+{c_1\over a_2}\chi_{_{A_1}}\right)-bx\left(1+{c_1\over 
a_2}\chi_{_{A_2}}\right)\right]^2+2\left[\left(1+{c_1\over a_2}\chi_{_{A_1}}
\right)^2+c^2y^2\left(1+{c_1\over a_2}\chi_{_{V}}\right)\right]^2, \non \\
&& x=\,{A_2^{BK^*}(m_{J/\psi}^2)\over A_1^{BK^*}(m_{J/\psi}^2)},~~~
y=\,{V^{BK^*}(m_{J/\psi}^2)\over A_1^{BK^*}(m_{J/\psi}^2)},~~~
z=\,{F_1^{BK}(m_{J/\psi}^2)\over A_1^{BK^*}(m_{J/\psi}^2)},
\en
and $\chi_{_{A_1}}$ denotes $\chi_{_{2,A_1^{BK^*}}}$ etc., for example [see
Eq.(5)],
\be
\chi_{_{A_1}}\equiv\chi_{_{2,A_1^{BK^*}}}
=\,{A_1^{(8)nf}(m_{J/\psi}^2)\over A_1^{BK^*}(m_{J/\psi}^2)}+
{a_2\over c_1}\,{A_1^{(0)nf}(m_{J/\psi}^2)\over A_1^{BK^*}(m_{J/\psi}^2)}.
\en
The analytic expressions of $a,~b$ and $c$ are given in \cite{Gour94,Alek1}. 
Numerically,
\be
a=3.164\,,~~~~b=1.304\,,~~~~c=0.435\,.
\en
The parity-odd ($P$-wave) transverse polarization measured in the 
transversity basis
\cite{Dighe}, which is more suitable for parity analysis, has the form
\be
|P|^2=\,{|A_\perp|^2\over |A_0|^2+|A_\| |^2+|A_\perp|^2},
\en
where $A_i$'s are defined by the orientation of the $J/\psi$ polarization
vector $\epsilon_{J/\psi}$ in the transversity basis \cite{CLEO96}: $A_0$ for 
$\epsilon_{J/\psi}$ parallel to $\hat{x}$, $A_\|$ for $\epsilon_{J/\psi}$
parallel to $\hat y$, and $A_\perp$ for $\epsilon_{J/\psi}$ parallel
to $\hat z$. Since $A_i$'s are related to the amplitudes $H_\lambda$
in the helicity basis via
\be
H_\pm={1\over\sqrt{2}}(A_\|\pm A_\perp),~~~~H_0=-A_0,
\en
we find
\be
|P|^2=\,2c^2y^2\left(1+{c_1\over a_2}\chi_{_V}\right)^2/\xi.
\en
A measurement of $|P|^2$ will thus provide information on the vector 
form factor.

   We would first like to see if the data of $\Gamma_L/\Gamma,~R$ and $|P|^2$ 
can be explained in the factorization approach. 
As we have stressed in passing, naive factorization does not work for the 
decay rate of color-suppressed
decay modes; nonfactorizable contributions
should always be included in order to describe the branching ratios
of $B\to J/\psi K(K^*)$. However, if $\chi_{_{F_1}}=\chi_{_{A_1}}=
\chi_{_{A_2}}=\chi_{_{V}}$, we have a new factorization scheme but
the predictions of $\Gamma_L/\Gamma,~R$ and $|P|^2$ in the naive 
factorization method remains intact since all nonfactorizable terms are 
canceled out in Eqs.(14) and (20). To proceed, we consider
several phenomenological models of form factors: (1) the Bauer-Stech-Wirbel
model (BSWI) \cite{BSW} in which $B\to K(K^*)$ form factors are first
evaluated at $q^2=0$ and then extrapolated to finite $q^2$ using a monopole
behavior for all form factors, (2) the modified BSW model (BSWII) 
\cite{BSWII}, which is
the same as BSWI except for a dipole $q^2$ dependence for form factors
$F_1,~A_0,~A_2$ and $V$, (3) the nonrelativistic quark model by Isgur,
Scora, Grinstein and Wise (ISGW) \cite{ISGW} with exponential $q^2$ 
dependence for all form factors, (4) the model of Casalbuoni {\it et al.}
and Deandrea {\it et al.} (CDDFGN) \cite{CDDFGN} in which form factors are 
first evaluated at $q^2=0$ using heavy meson effective chiral Lagrangians, 
which incorporate heavy
mesons, light pseudoscalar mesons and light vector mesons, and then 
extrapolated with monopole behavior. Several authors have derived the $B\to 
K(K^*)$ form factors from experimentally measured $D\to K(K^*)$ form factors
at $q^2=0$ using the Isgur-Wise scaling laws based on heavy quark symmetry 
\cite{IW}, which are allowed
to relate $B$ and $D$ form factors at $q^2$ near $q^2_{\rm max}$. The
$B\to K(K^*)$ form factors are calculated in \cite{Gour94} [IW(i)] by
assuming a monople dependence for all form factors, while they are computed
in \cite{CT} [IW(ii)] by advocating a monopole extrapolation for $F_1,~A_0,~
A_1$, a dipole behavior for $A_2,~V$, and an approximately constant for $F_0$. 
An ansatz proposed in \cite{Alek1}, which we call IW(iii), relies 
on ``soft'' Isgur-Wise scaling laws 
and different $q^2$ behavior of $A_1$ from $A_2,~V,~F_1$.

   In all above form-factor models, the $q^2$ dependence of form factors
is assumed to be governed by near pole (monopole or dipole) dominance. 
It is thus important to have a first-principles or model calculation
of the form-factor $q^2$ dependence. In principle,
QCD sum rules, lattice QCD simulations, and quark models allow one to 
compute form-factor
$q^2$ behavior. However, the analyses of the QCD sum rule yield some 
contradicting results. For example, while $A_1^{B\rho}$ is found to 
decrease from $q^2=0$ to $q^2=15\,{\rm GeV}^2$ in \cite{BBD}, such 
a phenomenon is not seen in \cite{ABS,YH}. Also the sum-rule results
become less reliable at large $q^2$ due to a large cancellation between
different terms. The present lattice QCD technique is not directly applicable
to the $B$ meson. Additional assumptions
on extrapolation from charm to bottom scales and from $q^2_{\rm max}$
to other $q^2$ have to be made.   As for the quark model, 
a consistent treatment
of the relativistic effects of the quark motion and spin in a bound state
is a main issue of the relativistic quark model. To our knowledge,
the light-front quark model \cite{Ter} is the only relativistic quark
model in which a consistent and fully relativistic treatment of quark spins
and the center-of-mass motion can be carried out.
A direct calculation
of $P\to P$ and $P\to V$ form factors at time-like momentum transfer in the
relativistic light-front (LF) quark model just became available recently 
\cite{CCH}. The contributions from valence-quark configuration to $B\to 
K(K^*)$ form factors are \cite{CCH}:
\footnote{In the light-front model calculations, the valence-quark 
contributions to form factors $A_0,~A_1,~V$ depend on the recoiling 
direction of the $K(K^*)$ relative to the $B$ meson \cite{CCH}. Thus the 
inclusion of the non-valence configuration arising from quark-pair
creation is in principle necessary in order to ensure that the physical
form factors are independent of the recoiling direction. Since the non-valence
contribution is most important only near zero recoil, we expect that
$Z$-graph effects on $B\to K(K^*)$ form factors at $q^2=m_{J/\psi}^2$
obtained in the ``+'' frame, where $K^*$ is moving in the $+z$ direction in 
the rest frame of the $B$ meson, are not
important. As noted in \cite{CCH}, the behavior of $V^{BK^*}$ in the
``+'' frame 
is peculiar in the sense that in general the form factor in the ``+'' frame
is larger than that in the ``$-$'' frame, so we have taken $V^{BK^*}$ from 
the ``$-$'' frame and other form factors from the ``+'' frame.}
\be
&& F_1^{BK}(m_{J/\psi}^2)=\,0.66,~~~~~A_1^{BK^*}(m_{J/\psi}^2)=\,0.37, \non \\
&& A_2^{BK^*}(m_{J/\psi}^2)=\,0.43,~~~~~V^{BK^*}(m_{J/\psi}^2)=\,0.50.
\en
We found that $F_1^{BK},~A_1^{BK^*},~A_2^{BK^*}$, 
$V^{BK^*}$ exhibit a dipole behavior, while
$A_1^{BK^*}$ shows a monopole dependence in the close 
vicinity of $q^2=0$ \cite{CCH}. Table I summerizes the predictions of 
$\Gamma_L/\Gamma,~R$ and $|P|^2$ in above-mentioned various form-factor 
models within the factorization approach by assuming the absence of inelastic
final-state interactions.

\begin{table}[t]
{{\small Table I. Predictions for form-factor ratios $x,~y,~z$
and for $\Gamma_L/\Gamma$, $R$ and $|P|^2$ in various form-factor models.}}
{
\begin{center}
\begin{tabular}{|c||c c c||c c c|} \hline
 & $x$ & $y$ & $z$ & $\Gamma_L/\Gamma$ & $R$ & $|P|^2$  \\  \hline
BSWI \cite{BSW} & 1.01 & 1.19 & 1.23 & 0.57 & 4.22 & 0.09  \\
BSWII \cite{BSWII} & 1.41 & 1.77 & 1.82 & 0.36 & 1.63 & 0.24 \\
ISGW \cite{ISGW} & 2.00 & 2.55 & 2.30 & 0.07 & 1.72 & 0.52 \\
CDDFGN \cite{CDDFGN} & 1.00 & 3.24 & 2.60 & 0.37 & 1.50 & 0.30 \\
IW(i) \cite{Gour94} & 0.98 & 2.56 & 1.74 & 0.45 & 2.89 & 0.31 \\
IW(ii) \cite{CT} & 0.88 & 1.77 & 2.05 & 0.56 & 1.84 & 0.16 \\
IW(iii) \cite{Alek1} & 1.08 & 2.16 & 1.86 & 0.45 & 2.15 & 0.26 \\
LF \cite{CCH} & 1.16 & 1.35 & 1.78 & 0.50 & 1.84 & 0.13 \\ \hline \hline
pre-1996 expt. & & & & $0.74\pm 0.07$ & $1.68\pm 0.33$ & -- \\ \hline
CDF \cite{CDF95,CDF96} & & & & $0.65\pm 0.11$ & $1.32\pm 0.28$ & -- \\
CLEO \cite{CLEO96} & & & & $0.52\pm 0.08$ & $1.36\pm 0.20$ & $0.16\pm 0.09$ 
\\ \hline
\end{tabular}
\end{center} }
\end{table}

   The pre-1996 values of $R$ and $\Gamma_L/\Gamma$ are given by
\be
R=\,1.68\pm 0.33\,,~~~~\Gamma_L/\Gamma=\,0.74\pm 0.07\,,
\en
where the former is the average value of ARGUS \cite{ARGUS} and CLEO
\cite{CLEO94}, and the latter is the combined average of ARGUS:
$\Gamma_L/\Gamma=0.97\pm 0.16\pm 0.15$ \cite{ARGUS}, CLEO:
$\Gamma_L/\Gamma=0.80\pm 0.08\pm 0.05$ \cite{CLEO94}, and CDF:
$\Gamma_L/\Gamma=0.65\pm 0.10\pm 0.04$ \cite{CDF95}. It is obvious from Table
I  that all the existing models fail to produce a large
longitudinal polarization fraction, whereas several models give
satisfactory results for the production ratio. 
Since the prediction of $\Gamma_L/\Gamma$ in  new factorization with
$\chi_{_{A_1}}=\chi_{_{A_2}}=\chi_{_{V}}$ is the same as that in naive
factorization, this means that {\it if the longitudinal polarization
fraction  is as large as $0.74\pm 0.07$, then
nonfactorizable terms should contribute differently to $S$-, $P$- and
$D$-wave amplitudes.} Consequently, the effective parameter $a_2^{\rm eff}$ 
cannot be defined for $B\to J/\psi K^*$ decays if $\Gamma_L/\Gamma$ is large.
In order to produce a large $\Gamma_L/\Gamma$,
various possibilities of nonfactorizable
contributions to $B\to J/\psi K^*$, e.g., $S$-wave dominance: $\chi_{_{A_1}}
\neq 0$, $\chi_{_{
A_2}}=\chi_{_{V}}=0$, have been explored in \cite{KS94,Cheng95,KA96}.

  The new results from CDF \cite{CDF96} on the branching-ratio measurements
of $B\to J/\psi K(K^*)$ decays give rise to
\be
R=\,1.32\pm 0.23\pm 0.16\,.
\en
Recently, CLEO II \cite{CLEO96} has completed the analysis of all data sample 
and come out with the results: \footnote{The previous CLEO result 
\cite{CLEO94}:
$\Gamma_L/\Gamma=0.80\pm 0.08\pm 0.05$ is based on a subset of the data used
in this complete analysis.} 
\be
R &=& 1.36\pm 0.17\pm 0.11\,,   \non \\
\Gamma_L/\Gamma &=& 0.52\pm 0.07\pm 0.04\,,   \\
|P|^2 &=& 0.16\pm 0.08 \pm 0.04\,.   \non
\en
Evidently, the new data of $\Gamma_L/\Gamma$ and $R$ tend to decrease. As
a consequence, the comparison between theory and experiment is turned 
the other way around: 
While several model predictions for $\Gamma_L/\Gamma$ are consistent with 
CLEO or CDF, almost all models (except for CDDFGN) fail to accommodate a 
small $R$. It
appears that IW(ii) and LF models are most close to the data: their
results for $\Gamma_L/\Gamma$ and $|P|^2$ are in agreement with experiment,
but the predicted $R$ is too large by 2 standard deviations compared to the 
experimental central value. Hence,
the factorization approach with $\chi_{_{A_1}}=\chi_{_{A_2}}=
\chi_{_{V}}$ can account for the longitudinal polarization fraction and 
parity-odd transverse polarization observed in $B\to J/\psi K^*$.
In order to explain the production ratio and $\Gamma_L/\Gamma$ simultaneously,
it is clear from Eq.(14) that we need
\footnote{For other form-factor models, one can always adjust small
nonfactorizable contributions (for example, $\chi_{_{A_1}}> \chi_{_{A_2}}
> \chi_{_{V}}$)
to accommodate the data, as illustrated in \cite{KS94,Cheng95,KA96}.
However, in practice, it is very difficult to pin down $\chi_{_{A_1}},~
\chi_{_{A_2}}$ and $\chi_{_{V}}$ separately if they are not the same. Our 
results (25) and (31) correspond to a minimal modification to the factorization
hypothesis.}
\be
\chi_{_{F_1}}>\chi_{_{A_1}}\sim\chi_{_{A_2}}\sim\chi_{_{V}},
\en
noting that $c_1/a_2>0$. 
Therefore, the data of $B\to J/\psi K(K^*)$ can be understood provided that
the nonfactorizable terms in $B\to J/\psi K^*$ are about the same in $S$-,
$P$-, and $D$ waves and that $\chi_{_{F_1}}$ in $B\to J/\psi K$ is slightly 
larger than that in $B\to J/\psi K^*$. The relativistic LF quark model
should be more reliable and trustworthy than the IW(ii) model since the 
$q^2$ behavior of form factors is directly calculated in the former.
We will thus confine ourselves to the LF model in ensuing discussion.

   To determine the magnitude of nonfactorizable terms we have to consider 
the branching ratios. The decay rates of $B\to J/\psi K$ and $B\to J/\psi K^*$
are given by
\be
\Gamma(B\to J/\psi K) &=& {p_c^3\over 4\pi}\,\left| a_2^{\rm eff}G_F V_{cs}
V_{cb}^*f_{J/\psi}F_1^{BK}(m_{J/\psi}^2)\right|^2   \non \\
&=& 1.102\times 10^{-14}{\rm GeV}\,|a_2^{\rm eff}F_1^{BK}(m_{J/\psi}^2)|^2,
\en
and
\be
\Gamma(B\to J/\psi K^*) &=& {p_c\over 16\pi m_B^2}\,\left| a_2^{\rm eff}G_F 
V_{cs}V_{cb}^*f_{J/\psi}m_{J/\psi}(m_B+m_{J/\psi})A_1^{BK^*}(m_{J/\psi}^2)
\right|^2 \non \\
&& \times \left[(a-bx)^2+2(1+c^2y^2)\right]    \\
&=& 1.19\times 10^{-14}{\rm GeV}\,|a_2^{\rm eff}A_1^{BK^*}(m_{J/\psi}^2)|^2
\left[(a-bx)^2+2(1+c^2y^2)\right],  \non
\en
where $p_c$ is the c.m. momentum, and uses of  
$ f_{J/\psi}=394$ MeV extracted from 
\footnote{Since we have taken into account the $q^2$ dependence of the 
fine-structure constant,
our values of $f_{J/\psi}$ and $f_{\psi'}$ given below
are slightly larger than the values cited in the literature.}
$J/\psi\to e^+e^-$ and $|V_{cb}|=0.038$ have been made. Fitting (26) and 
(27) to the  branching ratios
\be
{\cal B}(B^+\to J/\psi K^+) &=& \cases{ (1.08\pm 0.09\pm 0.09)\times 10^{-3};
\cr (1.01\pm 0.14)\times 10^{-3}, \cr}   \non \\
{\cal B}(B^0\to J/\psi K^0) &=& \cases{ (0.92^{+0.17}_{-0.15}\pm 0.08)\times 
10^{-3};  \cr (1.15\pm 0.23\pm 0.17)\times 10^{-3}, \cr}   \non \\
{\cal B}(B^+\to J/\psi K^{*+}) &=& \cases{ (1.41\pm 0.20\pm 0.24)\times 
10^{-3}; \cr (1.58\pm 0.47\pm 0.27)\times 10^{-3}, \cr}    \\
{\cal B}(B^0\to J/\psi K^{*0}) &=& \cases{ (1.32\pm 0.15\pm 0.17)\times 
10^{-3}; \cr (1.36\pm 0.27\pm 0.22)\times 10^{-3}, \rm}   \non
\en
where the upper entry refers to the CLEO data \cite{CLEO96} and the lower
entry (except for ${\cal B}(B^+\to J/\psi K^+)$ which comes from the PDG 
\cite{PDG}) to CDF data \cite{CDF96}, we find averagely
\be
a_2^{\rm eff}(B\to J/\psi K)=\,0.30\,,~~~~~a_2^{\rm eff}(B\to J/\psi K^*)=\,
0.26\,,
\en
for the lifetimes \cite{PDG}
\be
\tau(B^0)=\,1.56\times 10^{-12}s,~~~~~\tau(B^\pm)=\,1.62\times 10^{-12}s,
\en
and the form factors (21).
Using $c_1(m_b)=1.12$ and $c_2(m_b)=-0.27$ for $\Lambda^{(5)}_{\rm QCD}=200$
MeV, we obtain
\be
\chi_{_{F_1}}=\,0.17\,,~~~~~\chi_{_{A_1}}\sim\chi_{_{A_2}}\sim\chi_{_{V}}=
0.14\,.
\en
Since $|a_2^{\rm eff}(B\to J/\psi K^{(*)})|$ is numerically close to 
$|c_2(m_b)|$, one may tempt to argue that the large-$N_c$ approach works 
also for $B$ decays. However, this possibilty is ruled out by the observed 
constructive
interference in charged $B^\pm \to D^{(*)}\pi(\rho)$ decays \cite{CLEO94},
which implies a positive ratio $a_2/a_1$. Therefore,
the sign of $a_2^{\rm eff}$ in (29) and $\chi_2$ in (31) is fixed to be
positive, in dramatic contrast to the charm case.
The result (31) supports the conjecture (13), but the question of why the
nonfactorizable term, which lies in the range $-0.60<\chi_2<
-1/3$ for charm decay \cite{Cheng95}, 
 becomes positive in $B$ decay remains mysterious and baffling. 
(For a recent attempt of understanding the positive
sign of $a_2^{\rm eff}$ and $\chi_2$ in $B$ decay, see \cite{Li}.)

   With the result (29) or (31) we are ready to discuss
$B\to \psi' K(K^*)$ decays where $\psi'\equiv\psi(2S)$. The
analytic expressions for analogus $\Gamma'_L/\Gamma',~R'$ and $|P'|^2$
are the same as (14), (15) and (20) except that the coefficient 1.08 in (14)
is replaced by 2.45 and that the unprimed quantities $a,~b,~c,~x,~y,~z$
are replaced by primed ones:
\be
&& a'=\,2.051,~~~~b'=\,0.733,~~~~c'=\,0.356,   \non \\
&& x'=\,{A_2^{BK^*}(m_{\psi'}^2)\over A_1^{BK^*}(m_{\psi'}^2)},~~~
y'=\,{V^{BK^*}(m_{\psi'}^2)\over A_1^{BK^*}(m_{\psi'}^2)},~~~
z'=\,{F_1^{BK}(m_{\psi'}^2)\over A_1^{BK^*}(m_{\psi'}^2)}.
\en
We find in the LF model that
\be
F_1^{BK}(m_{\psi'}^2)=\,0.92,~~A_1^{BK^*}(m_{\psi'}^2)=\,0.44,~~
A_2^{BK^*}(m_{\psi'}^2)=\,0.55,~~V^{BK^*}(m_{\psi'}^2)=\,0.66.
\en
Using $f_{\psi'}=293$ MeV determined from $\psi'\to e^+e^-$, and assuming
$a_2^{\rm eff}(B\to \psi'K)\sim a_2^{\rm eff}(B\to J/\psi K)$ and
$a_2^{\rm eff}(B\to \psi'K^*)\sim a_2^{\rm eff}(B\to J/\psi K^*)$, we obtain
\be
&& {\cal B}(B^0\to \psi' K^0)=\,0.50\times 10^{-3},~~~~{\cal B}(B^+\to\psi' 
K^+)=\,0.52 \times 10^{-3},   \non \\
&& {\cal B}(B^0\to \psi' K^{*0})=\,0.76\times 10^{-3},~~~~{\cal B}(B^+\to\psi' 
K^{*+})=\,0.79 \times 10^{-3},
\en
and
\be
{\Gamma'_L/\Gamma'}=\,0.33\,,~~~~~R'=\,1.57\,,~~~~~|P'|^2=\,0.15\,.
\en
Our results for branching ratios agree with the new CDF
measurements \cite{CDF96}:
\be
{\cal B}(B^+\to\psi' K^+) &=& (0.68\pm 0.10\pm 0.14)\times 10^{-3}, \non \\
{\cal B}(B^0\to \psi'K^{*0}) &=& (0.90\pm 0.21\pm 0.20)\times 10^{-3}.
\en
Note that our prediction of $\Gamma'_L/\Gamma'$ is quite different from
the predicted range $0.50\lsim \Gamma'_L/\Gamma'\leq 0.67$
given in \cite{KS95b}.

    Finally we trun to the decay mode $B_s^0\to J/\psi\phi$. Following 
\cite{CCH}, the relevant form factors are calculated in the LF model to be
\be
A_1^{B_s\phi}(m_{J/\psi}^2)=0.35\,,~~~A_2^{B_s\phi}(m_{J/\psi}^2)=0.39\,,~~~
V^{B_s\phi}(m_{J/\psi}^2)=0.49\,.
\en
A straightforward calculation yields
\be
{\Gamma_L/ \Gamma}(B_s\to J/\psi\phi)=0.51\,,~~~~~
|P|^2(B_s\to J/\psi\phi)=0.13\,,
\en
and
\be
{\cal B}(B_s\to J/\psi\phi)=\,1.29\times 10^{-3}
\en
for $a_2^{\rm eff}\sim 0.25$. Experimentally, longitiudinal polarization
fraction and the branching ratio have been measured by CDF:
\be
\Gamma_L/\Gamma=0.56\pm 0.21^{+0.02}_{-0.04}~~\cite{CDF95},~~~~
{\cal B}(B_s\to J/\psi \phi)=(0.93\pm 0.28\pm 0.17)\times 10^{-3}~~
\cite{CDF96}.
\en

   To summarize, naive factorization or new factorization with universal
nonfactorizable terms cannot explain the data of $\Gamma_L/\Gamma,~R$ and
$|P|^2$ simultaneously in all existing form-factor models. However, we have 
shown that a minimal
modification to the factorization hypothesis can accommodate the
recently available data from CDF and CLEO: the nonfactorizable
term $\chi_{_{F_1}}$ in $B\to J/\psi K$ should be slightly larger than
$\chi_{_{A_1}}\sim\chi_{_{A_2}}\sim\chi_{_{V}}$ in $B\to J/\psi K^*$
decays. When generalized to $B\to \psi(2S)K(K^*)$ and $B_s\to J/\psi\phi$, our
predictions agree with presently available measurements. Generally speaking,
the bulk of exclusive decays of heavy mesons can be grossly accounted for
by the factorization approach with universal nonfactorizable
contribution $\chi\lsim -{1\over 3}$ for charm decays and with $\chi\sim 0.15$ 
for bottom decays; that is, the new factorization scheme is very different
for charm and bottom decays.
Just as the charm case, the data of $B\to J/\psi K(K^*)$
start to reveal departures from the factorization postulation: the
nonfactorizable term $\chi$ in $B$ decay is also process dependent and not
universal.

\vskip 2.0cm
\centerline{\bf ACKNOWLEDGMENTS}
\vskip 0.3 cm
   I wish to thank C.W. Hwang for using the framework of \cite{CCH} to 
compute the form factors relevant to this manuscript.
    This work was supported in part by the National Science Council of ROC
under Contract No. NSC86-2112-M-001-020.

\renewcommand{\baselinestretch}{1.1}
\newcommand{\bi}{\bibitem}

\newpage
\begin{thebibliography}{99}
%

\bi{Gour94}
M. Gourdin, A.N. Kamal, and X.Y. Pham, \prl {\bf 73}, 3355 (1994).

\bi{Alek1} R. Aleksan, A. Le Yaouanc, L. Oliver, O. P\`ene, and J.-C.
Raynal, \pr {\bf D51}, 6235 (1995). 

\bi{CDF96}
CDF Collaboration, F. Abe {\it et al.,} FERMILAB-PUB-96/119-E; 
FERMILAB-Conf-96/160-E (1996).

\bi{CLEO96}
CLEO Collaboration, D.M. Asner {\it et al.,} CLEO-CONF 96-24 (1996).

\bibitem{Cheng89}
For a review, see e.g., H.Y. Cheng, {\sl Int. J. Mod. Phys.} {\bf A4}, 495 
(1989).


\bibitem{PDG}
Particle Data Group, \pr {\bf D54}, 1 (1996).
%
\bibitem{Cheng94}
H.Y. Cheng, \pl {\bf B335}, 428 (1994).
%
%
\bibitem{Soares}
J.M. Soares, \pr {\bf D51}, 3518 (1995).

\bi{KS95}
A.N. Kamal and A.B. Santra, Alberta Thy-08-95 (1995).

%
\bibitem{KS94}
A.N. Kamal and A.B. Santra, Alberta Thy-31-94 (1994).

\bibitem{Fuk}
M. Fukugita, T. Inami, N. Sakai, and S. Yazaki, \pl {\bf 72B}, 237 (1977); 
D. Tadi\'c and J. Trampeti\'c, \pl {\bf 114B}, 179 (1982); M. Bauer and B. 
Stech, \pl {\bf 152B}, 380 (1985).

\bibitem{Buras}
A.J. Buras, J.-M. G\'erard, and R. R\"uckl, \np {\bf B268}, 16 (1986).

\bibitem{Cheng95}
H.Y. Cheng, \zp {\bf C69}, 647 (1996).

\bi{Dighe} A.S. Dighe, I. Dunietz, H.J. Lipkin, and J.L. Rosner, \pl {\bf
B369}, 144 (1996).

\bibitem{BSW} M. Bauer, B. Stech, and M. Wirbel, \zp {\bf C34}, 103 (1987).

%
\bibitem{BSWII}
M. Neubert, V. Riekert, Q.P. Xu, and B. Stech, in {\it Heavy Flavors}, 
edited by A.J. Buras and H. Lindner (World Scientific, Singapore, 1992).

\bi{ISGW} N. Isgur, D. Scora, B. Grinstein, and M. Wise, \pr {\bf D39}, 799
(1989).

\bi{CDDFGN} R. Casalbuoni, A. Deandrea, N. Di Bartolomeo, R. Gatto, F. 
Feruglio,
and G. Nardulli, \pl {\bf B299}, 139 (1993); A. Deandrea, N. Di Bartolomeo,
R. Gatto, and G.Nardulli, \pl {\bf B318}, 549 (1993).

\bi{IW} N. Isgur and M. Wise, \pr {\bf D42}, 2388 (1990).

\bi{CT} H.Y. Cheng and B. Tseng, \pr {\bf D51}, 6259 (1995).

\bi{BBD} P. Ball, V.M. Braun, and H.G. Dosh, \pr {\bf D44}, 3567 (1991);
P. Ball, \pr {\bf D48}, 3190 (1993).

\bi{ABS} A. Ali, V.M. Braum, and H. Simma, \zp {\bf C63}, 437 (1994).

\bi{YH} K.C. Yang and W-Y.P. Hwang, NUTHU-94-17, to appear in {\sl Z. Phys.}
(1996).

\bi{Ter} M.V. Terent'ev, {\sl Sov. J. Phys.} {\bf 24}, 106 (1976); V.B.
Berestetsky and M.V. Terent'ev, {\sl ibid.} {\bf 24}, 547 (1976); 
{\bf 25}, 347 (1977); P.L. Chung, F. Coester, and W.N. Polyzou,
\pl {\bf B205}, 545 (1988).

\bi{CCH} H.Y. Cheng, C.Y. Cheung, and C.W. Hwang, IP-ASTP-04-96 
[hep-ph/9607332].

\bibitem{ARGUS}
ARGUS Collaboration, H. Albrecht {\it et al.,} \pl {\bf B340}, 217 (1994).

\bibitem{CLEO94}
CLEO Collaboration, M.S. Alam {\it et al.,} \pr {\bf D50}, 43 (1994).

\bi{CDF95} CDF Collaboration, F. Abe {\it et al.,} \prl {\bf 75}, 3068 (1995).

\bi{KA96} A.N. Kamal and F.M. Al-Shamali, Alberta Thy-12-96 [hep-ph/9605293].

\bi{Li} C-H. V. Chang and H-n. Li, CCUTH-96-03 [hep-ph/9607214].

\bi{KS95b} A.N. Kamal and A.B. Santra, \pr {\bf D51}, 1415 (1995).

\end{thebibliography}
\end{document}